\documentclass{www2012-accepted}

\usepackage[utf8]{inputenc}
\usepackage{graphicx}

\usepackage{booktabs}
\usepackage{longtable}
\usepackage{rotating}

\begin{document}

\clubpenalty=10000 
\widowpenalty = 10000

\title{Dynamical Classes of Collective Attention in Twitter}

\numberofauthors{4}
\author{
\alignauthor
Janette Lehmann\\
       \affaddr{Web Research Group, Universitat Pompeu Fabra}\\
       \affaddr{Barcelona, Spain}\\
       \email{janette.lehmann@gmx.de}
\alignauthor
Bruno Gon\c calves\\
       \affaddr{College of Computer and Information Sciences}\\
       \affaddr{Northeastern University}\\
       \email{b.goncalves@neu.edu}
\and
\alignauthor 
Jos\'e J. Ramasco\\
       \affaddr{IFISC (CSIC-UIB)}\\
       \affaddr{Palma de Mallorca, Spain}\\
       \email{jramasco@ifisc.uib.es}
\alignauthor Ciro Cattuto\\
       \affaddr{ISI Foundation}\\
       \affaddr{Torino, Italy}\\
       \email{ciro.cattuto@isi.it}
}


\maketitle
\begin{abstract}
Micro-blogging systems such as Twitter expose digital traces of social discourse with an unprecedented degree of resolution of individual behaviors. They offer an opportunity to investigate how a large-scale social system responds to exogenous or endogenous stimuli, and to disentangle the temporal, spatial and topical aspects of users' activity. Here we focus on spikes of collective attention in Twitter, and specifically on peaks in the popularity of hashtags. Users employ hashtags as a form of social annotation, to define a shared context for a specific event, topic, or meme. We analyze a large-scale record of Twitter activity and find that the evolution of hashtag popularity over time defines discrete classes of hashtags. We link these dynamical classes to the events the hashtags represent and use text mining techniques to provide a semantic characterization of the hashtag classes. Moreover, we track the propagation of hashtags in the Twitter social network and find that epidemic spreading plays a minor role in hashtag popularity, which is mostly driven by exogenous factors.
\end{abstract}




\section{Introduction}
\label{sec:intro}

Popularity plays a major role in the dynamics of online systems. Public attention can suddenly concentrate on a Web page or application~\cite{vlachos04,wu07,adar07,onnela09,goel10,ratkiewicz10b}, a Youtube video~\cite{crane08,figueiredo11,naaman11}, a trending topic in Twitter~\cite{kwak10,asur11,yang11}, or on a story in the news media~\cite{leskovec09}, sometimes even in absence of an apparent reason. Typically, after an initial increase of attention, the focus will move elsewhere leaving as a trace a characteristic activity profile. Such popularity peaks are not only of great relevance for the monetization of  online content, but also pose scientific challenges related to understanding the mechanisms ruling their dynamics~\cite{wu07,crane08,onnela09,ratkiewicz10,asur11}.
In particular, specific features of the popular item under consideration can now be related to its activity profile by means of semantic analysis and natural language processing of the messages exchanged by the users~\cite{adar07,huang10,wu11}.

Here we use data from the Twitter micro-blogging system to investigate the relation between activity profiles over time and content. There are several reasons for selecting Twitter: It is one of the most popular online social networks, part of its message stream is programmatically accessible to the public~\cite{API}, and the content of the messages is short, making it amenable to automated processing. Twitter is used as an hybrid between a communication media and an online social network~\cite{kwak10,wu11} and hosts real-time discussion of current topics of popular interest. We take advantage of the practice introduced by Twitter users of attaching ``hashtags'' to their messages as a way of explicitly marking the relevant topics. Twitter has incentivated this practice by supporting hashtags in their Web interface and in their programmatic API, turning them into lightweight social annotations of the information streams users consume. Here we focus our analysis on those hashtags that exhibited a popularity peak during our observation period, and systematically analyze the corresponding messages (``tweets'') by grounding the words they contain in a semantic lexicon.

This paper is structured as follows: Section~\ref{sec:background} reviews the literature on Twitter and in particular the literature on temporal patterns of Twitter activity. Section~\ref{sec:data} describes the Twitter dataset we used and the techniques we applied to select popular hashtags and their usage patterns. In Section~\ref{sec:classes} we identify dynamical classes of hashtag usage and relate them to the semantics of the corresponding tweets. In Section~\ref{sec:spreading} we relate the same dynamical classes to the spreading properties of hashtags over the underlying social network. Section~\ref{sec:discussion} summarizes our findings and points to applications and further research directions.

\section{Related Work}
\label{sec:background}

Several aspects of Twitter have been extensively investigated in the literature, including its network topology~\cite{java07,krishnamurthy08,huberman08}, the relations and types of messages between users~\cite{honeycutt09,cha10}, the internal information propagation~\cite{galuba10,Lerman_Ghosh_2010,moreno2011}, the credibility of information~\cite{mendoza10,castillo11}, and even its potential as an indicator of the state of mind of a population~\cite{jansen09,ratkiewicz11-1,ratkiewicz11-2,dodds11,DBLP:journals/computer/BollenM11,moreno2011}.

The possibility that popular trends or hashtags could be classified in groups have been discussed in Refs.~\cite{kwak10,wu11,yang11}, and the effect of semantic differences on the persistence of a hashtag have also been considered~\cite{romero11}. The shape of peaks in popularity profiles has been used to classify the events in groups~\cite{crane08,kwak10,laniado10,yang11}. The hypothesis that both the increase and decrease of public attention follow a power-law-like functional shape whose exponents define universality classes, in parallel to what occurs with phase transitions in critical phenomena, has been explored~\cite{crane08}. This approach, however, is difficult to apply to Twitter: the fast timescales involved and the highly reactive nature of Twitter make the time series very noisy and pose the challenge of characterizing activity dynamics in a way which is both robust and scalable.

The  causes that underlie the existence of distinct classes of popularity are thought to be a combination of all the mechanisms that drive public attention. News regarding a popular item can propagate either over the social network of the users of a given system -- a so-called endogenous process --
or it can be injected through mass media (exogenous driving). The duality between exogenous and endogenous information propagation has permeated the analysis of popularity in several recent studies~\cite{crane08,kwak10,figueiredo11,naaman11}, even though it is not always clear how to distinguish between them based solely on the shape of the respective popularity profiles~\cite{naaman11}.

\section{Data}
\label{sec:data}

Our dataset comprises about $130$ million Twitter messages or \textit{tweets} posted between November $20$, $2008$ and May $27$, $2009$. The data were collected at Indiana University thanks to their temporary privileged access to the Twitter data stream~\cite{goncalves11}. Each tweet includes textual content, an author, the time at which it was posted, whether or not it was in reply to another tweet, and additional metadata. The collected tweets come from about $6.1$ million unique user accounts.

In order to build a representation of the social network over which hashtag diffusion takes place, we queried the Twitter REST API for the complete list of followers and friends of $3.5$ million users. We collected neighbor information for $2.7$ million of them, the discrepancy being accounted for by users with a private profile. Using this information we constructed a directed follower network, where each edge takes on the direction in which information flows: if user A follows user B, the respective social link points from B to A, as A can see B's status updates.

\subsection{Hashtags Selection}

For the identification of topics, we extracted all the hashtags contained in the Twitter messages (by matching the tweet content to the pattern \textquotedblleft \#[a-zA-Z0-9\_]* \textquotedblright). Our dataset includes about $400,000$ distinct hashtags (see Table~\ref{tab:data}). We selected the most popular topics by restricting our data to the hashtags used by at least $500$ distinct users and to the messages containing at least one of such hashtags. Based on this selection, we used for the following analysis about $1.7$ million tweets and $402$ popular hashtags.
\begin{table}[h]
	\centering
		\begin{tabular}{p{150pt}r}
			\hline
			total number of tweets & $131,737,688$ \\
			total number of tweets with hashtags & $4,292,929$ \\
			total number of hashtags &	$408,254$ \\
			total number of users & $6,477,072$ \\
			average number of tweets per user & $20.34$ \\
			
			\hline
		\end{tabular}
\caption{General statistics about the dataset
\label{tab:data}
}
\end{table}

\subsection{Activity Peak Detection}

Like most systems driven by human actions, Twitter exhibits bursty activity, circadian rhythms, and in general the full temporal complexity of a large-scale social aggregate. Because of this, there is no single natural scale for investigating its temporal behavior, and the choice of a time scale is not neutral with respect to the phenomena one can study at that scale. Here we choose to investigate activity at the scale of days, i.e., we do not study human dynamics at the level of minutes and seconds, nor phenomena driven by the circadian cycle, nor slower trends that develop over several weeks of months. We analyze daily activity levels, and focus on events that are meaningful at that scale, such as the wait for a scheduled social event.

At the daily scale the popularity profile of hashtags can look very different. On visual inspection the individual temporal profiles of hashtag usage display behaviors that typically fall into one of the following three categories: continuous activity, periodic activity, or activity concentrated around an isolated peak.

Continuous-activity profiles are those for which a rather constant level of daily activity is maintained by the user community  (e.g., \texttt{music}). Hashtags with periodic activity profiles display series of spikes spaced by one or more weeks, or months  (e.g., \texttt{followfriday}). Finally, activity profiles with an isolated peak are characteristic of hashtags associated with a unique event to which a user community pays attention for a limited span of time (e.g., \texttt{oscars}). In the following we will concentrate on this class of hashtags.

To identify activity peaks, for every hashtag $H$ we compute the time series of daily activity, where the activity $n_H(i)$ on day $i$ is defined as the number of tweets containing $H$. In the following we will write $n(i)$ to indicate the activity level of a generic hashtag. We use a sliding window of $2L+1$ days ($L=30$) centered on day $i_0$,
$T = [ n(i_0-L), n(i_0-L+1), \ldots , n(i_0-1), n(i_0), n(i_0+1), ... , n(i_0+L-1), n(i_0+L) ]$,
 and let $i_0$ slide along the activity time series for the hashtag. Within this window we evaluate the baseline hashtag activity as the median $n_b$ of $T$. Then, we define the outlier fraction $p(i_0)$ of the central day $i_0$ as the relative difference of the hashtag activity $n(i_0)$ with respect to the median baseline $n_b$: $p(i_0) = [n(i_0) - n_b] / \max(n_b, n_{\mbox{\tiny min}})$. Here $n_{\mbox{\tiny min}} = 10$ is a mininum activity level used to regularize the definition of $p(i_0)$ for low activity values. We say that there is an activity peak at $i_0$ if $p(i_0) > p_t$, where $p_t$ is an arbitrary threshold value that in the following we set as $p_t = 10$. We checked that different values of the threshold do not change significantly our results, and that the same peaks can be identified by using different peak-detection techniques.

\begin{figure*}
\centering
\includegraphics[width=0.9\textwidth]{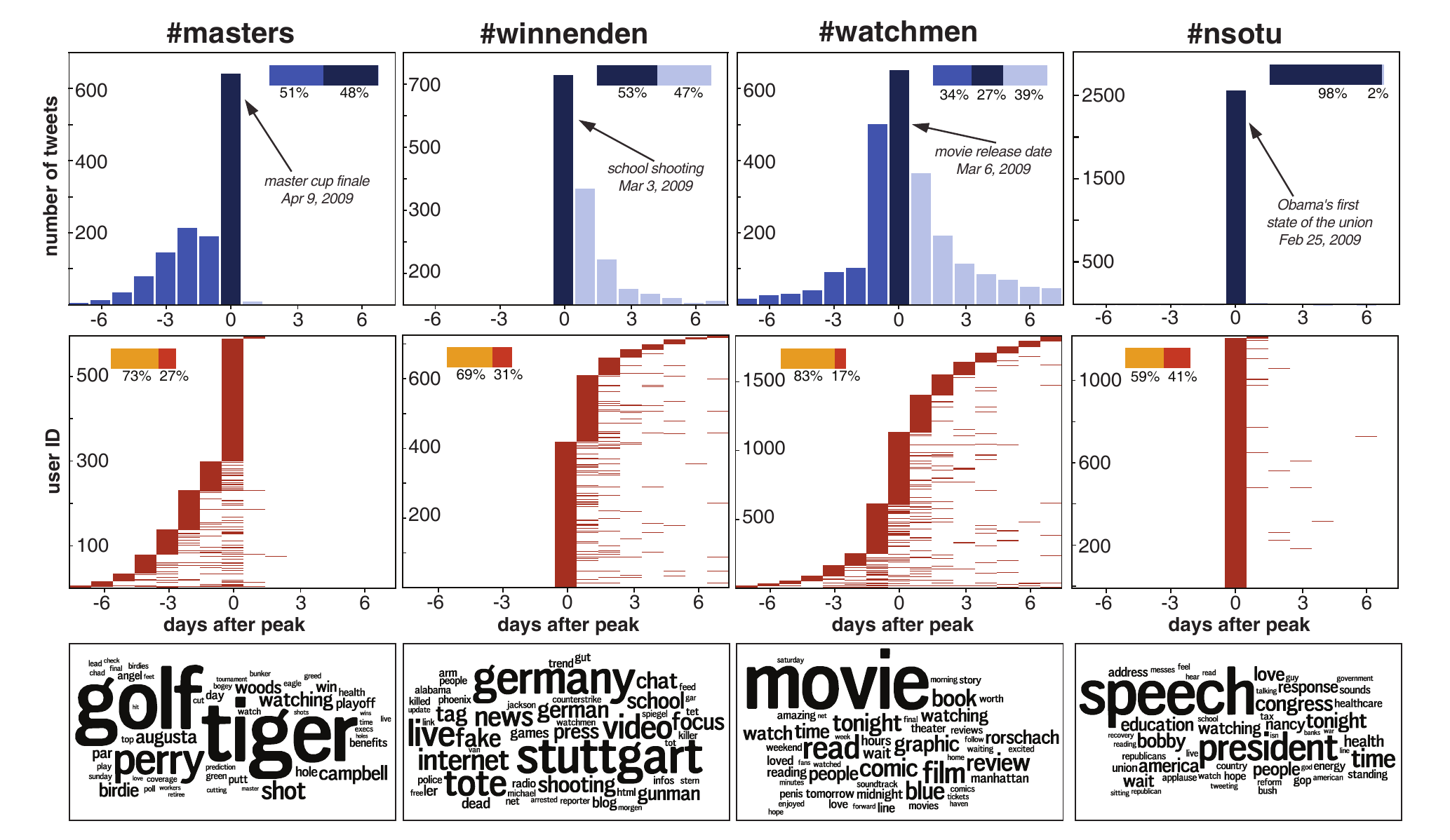}
\caption{Activity associated with four hashtags that exhibit a popularity peak: daily activity over time (top row), individual user activity (middle) row, and word clouds of tweet content (bottom row). 
\label{activity}
}
\end{figure*}

Of course it may happen that for a given hashtag $H$ the time series $n_H(i)$ exhibits more than one peak. Since we are interested in isolated popularity bursts, we ignore all peaks that are separated from other peaks by less than one week. Finally, for every hashtag we select the peak (if any) with the highest $p(i_0)$ and we offset the day index so that for all hashtags the activity peak occurs on day $0$, as shown in Fig.~\ref{activity}.
Using this method we select $115$ peaks of daily hashtag activity: the corresponding hashtags are listed in Appendix~\ref{appendix-usage}, together with manual annotations about their meaning and a coarse classification.

\subsection{Semantic Grounding}
\label{semanticanalysis}

To correlate the temporal activity patterns with content, we perform a simple semantic grounding of the tweets by using the WordNet~\cite{wordnet} semantic lexicon. For each tweet, we pre-process the text by removing user mentions (\texttt{@username}), hashtags, URLs and a standard set of English stop words. Then, for each word we perform stemming (with the standard Porter algorithm), lemmatization, and we finally attempt to look up in WordNet the corresponding \text{synset} (i.e., the basic node of the WordNet lexicon, a set of synonyms that refer to a single concept). From now on we will refer to WordNet synsets as \textit{concepts}. Words for which no concept can be looked up in WordNet are ignored.
%
If few or no terms are successfully looked up in WordNet as English words, we attempt to identify the tweet language: we run the TextCat~\cite{textcat} language categorization algorithm on the text and we discard the tweet if English is not included in the top $10$ most likely languages identified by TextCat.

Overall, the above analysis identifies about $18,000$ distinct concepts that are associated with the hashtags under study. 

\subsection{Social attention and popular hashtags}
\label{poptags}

Typical examples of the activity profiles for the selected hashtags  are shown in Figure~\ref{activity}. The curves are centered around the day on which the popularity reaches its maximum (day $0$).  The displayed time window spans one week before and after the peak.
In the top plots of Figure~\ref{activity} the activity of four sample hashtags is reported as a function of time in days after the peak. The bars on the top right display the percentage of activity before, at and after the peak. The four hashtags exhibit different behaviors in terms of approach to the peak (dark blue bars) and relaxation after the peak (light blue bars). The hashtag \texttt{masters} exhibits an anticipatory pattern, with a gradual build-up of activity before the peak. The hashtag \texttt{winnenden}, conversely, corresponds to an unexpected event, with a sudden onset of activity followed by a gradual relaxation. The hashtag \texttt{watchmen} displays both a gradual build-up of attention and a gradual relaxation after the peak. Finally, the hashtag \texttt{nsotu} concentrates almost all of its activity during the single day of the peak.
In the middle plot row we show the activity of individual users as a function of time. Users who have posted the hashtag at least once (within the observation interval) are ranked according to the time of first usage of the hashtag (rank along the ordinate axis): early adopters lie at the bottom and late adopters are at the top. For each user, colored segments mark the times at which the hashtag under consideration was used. The inset bar plots show the fraction of users who used the hashtag more than once during the selected time window.
%
Finally, in the bottom plot row we visualize the content of tweets as word clouds. Each word cloud contains the $50$ most frequent words, with font sizes proportional to word frequencies.
The patterns displayed by these hashtags are representatives of the four classes of activity peaks found in our analysis.

\section{Classes of Popular Hashtags}
\label{sec:classes}

\begin{figure}
\centering
\includegraphics[width=0.85\columnwidth]{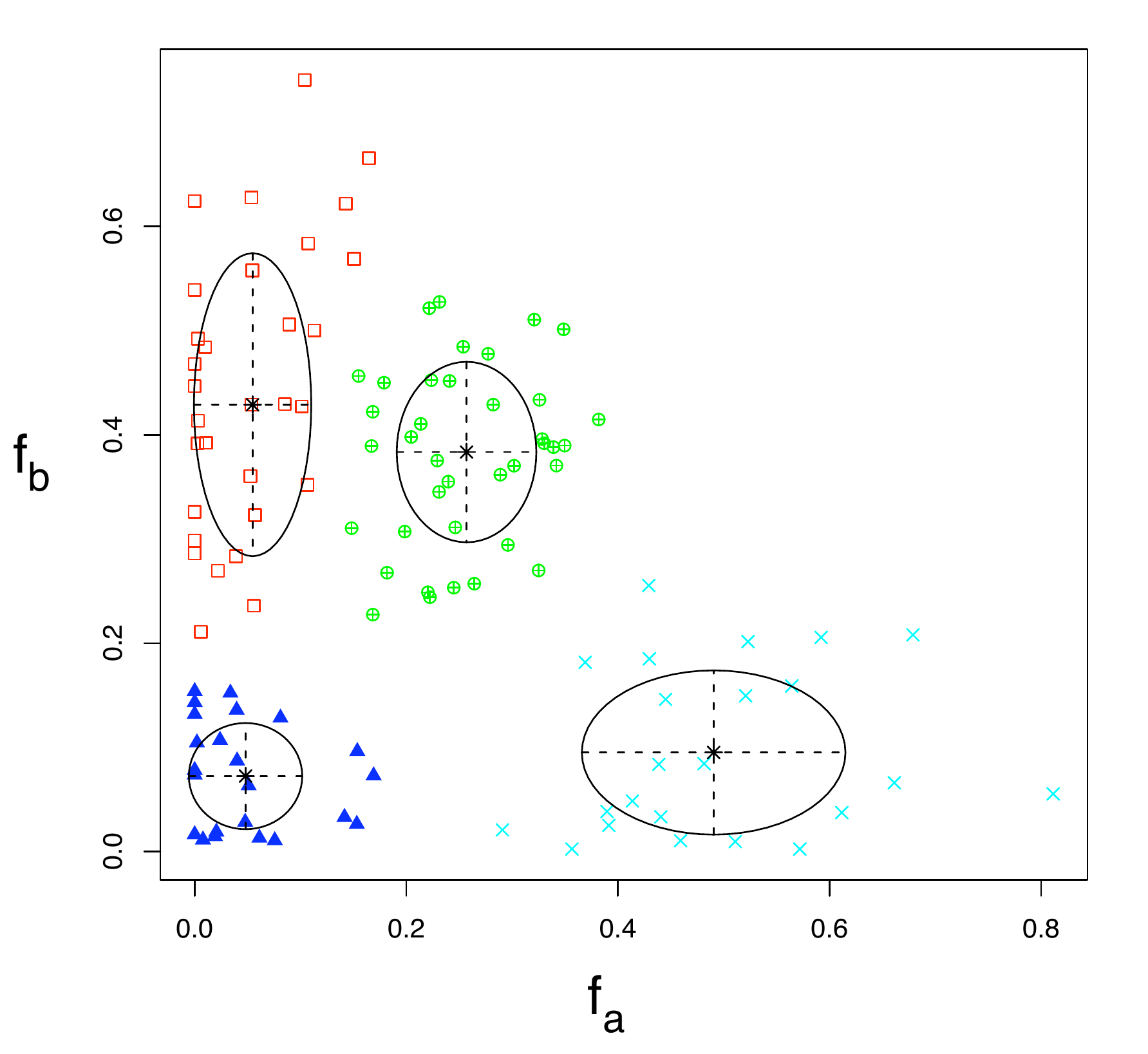}
\caption{
Mixture Gaussian model learned by using the Mclust implementation of the Expectation Maximization algorithm. The individual components have variable variance along both the $f_a$ and $f_b$ axes (VVI model of the Mclust implementation). 
\label{em-clusters}
}
\end{figure} 

\begin{figure*}[t]
\centering
\includegraphics[width=1.3\columnwidth]{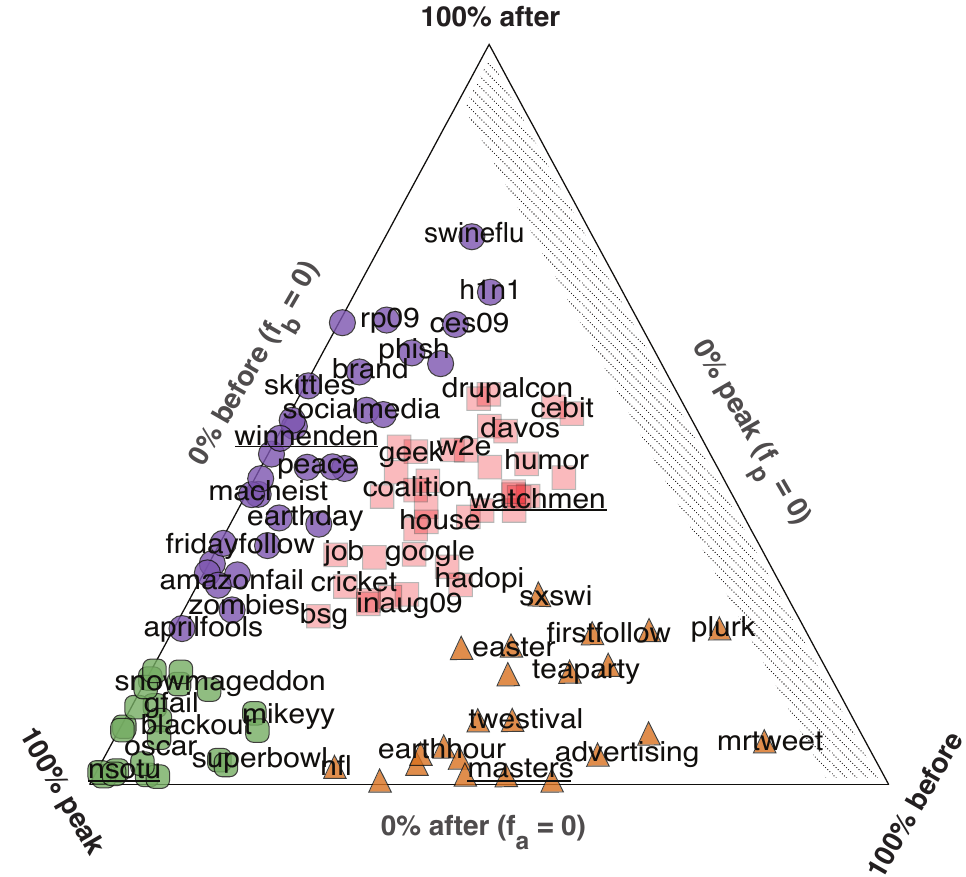}
\caption{The four hashtag clusters in the $(f_b,f_p,f_a)$ simplex. Orange triangles: activity concentrated before an event.~
Purple circles: activity concentrated after an event.~ 
Red squares:  symmetric activity.~ 
Green round squares: activity concentrated on the day of an event. 
The hashtags of Fig.~\ref{activity} are underlined.
\label{simplex}}
\end{figure*}

The possibility of classifying online popularity peaks in a few discrete classes has been discussed in the literature~\cite{crane08,kwak10,figueiredo11,yang11,naaman11}. Typically the classification is done according to the different shapes or functional forms of the increasing and decreasing parts of the popularity profiles. The origin of these few classes has been linked in the literature to two mechanisms that, to some extent, are present in most online social systems: endogenous propagation of information over the social network, and the injection into the system of information from exogenous online or offline sources. This scenario was tested for the evolution of popularity of YouTube videos~\cite{crane08,figueiredo11} and has also been discussed for trending topics or memes in Twitter~\cite{kwak10,yang11,naaman11}. The lack of a clear distinction between endogenous and exogenous information flow in Twitter means that the number of classes, the possible functional shapes of the popularity profiles, and even the importance of the endogenous/exogenous distinction are all far from clear~\cite{naaman11}.

Here we take a different approach and attempt to simplify the possible scenarios by shifting emphasis from the detailed time series of popularity to coarse-grained information on the balance of activity before, during, and after the popularity peak. To achieve this, for each hashtag exhibiting a popularity peak we summarize the hashtag usage timeline with the triple $(f_b, f_p, f_a)$ of the fractions of tweets posted before ($f_b$), during ($f_p$) and after the peak ($f_a$). By definition these fractions satisfy $f_b+f_p+f_a = 1$. We restrict the computation to a two-week period centered on the peak time, as shown in the examples of Fig.~\ref{activity}.

\subsection{Identifying Classes}

We identify hashtag clusters in the $(f_b,f_a)$ space of independent parameters using a standard implementation of the Expectation Maximization (EM) algorithm~\cite{mclust, fraley02} to learn an optimal Gaussian mixture model. The number of components (clusters) of the mixture is set by using the Bayesian Information Criterion, as well as by means of a $10$-fold cross-validation, yielding in both cases the $4$ clusters shown in Fig~\ref{em-clusters}. The clusters are robust with respect to the initial conditions and parameters of the EM algorithms (provided that care is taken to deal with the points on the $f_b = 0$ axis): $77\%$ of the hashtags have a classification accuracy below $5\%$, and only $6\%$ of them have a classification accuracy in excess of $20\%$.

Figure~\ref{simplex} shows the identified clusters in the $3$-simplex $(f_b,f_p,f_a)$. The marker representing each of the $115$ selected hashtag is colored and shaped according to the group it has been classified into. The hatched area is the parametric space excluded by the constraint that hashtags should have a peak-day activity of at least $10$ times the baseline daily activity (i.e., the excluded parametric space is due to our selection of hashtags that exhibit a peak in their activity timeline). The four groups of Fig.~\ref{simplex} correspond to different temporal patterns of collective attention, as illustrated below in relation to the hashtags of Fig.~\ref{activity}.
\begin{itemize}
	\item Activity concentrated \textit{before and during} the peak (orange triangles).
	 These hashtags correspond, by definition, to anticipatory behavior, with users posting increasing amount of content as the date of the event approaches, followed by a sharp drop in attention right after the event. See for example the hashtag \texttt{\#masters} (underlined in the figure) which was used to discuss the $2009$ Golf Masters.
	\item Activity concentrated \textit{during and after} the peak (purple circles). In this class we find hashtags indicating unexpected events that make an impact, such as the \texttt{\#winnenden} school shooting. The sudden onset of activity is a reaction to the unexpected event, and it is likely to be driven by exogenous sources such as communication in mass media.%
	\item Activity concentrated \textit{symmetrically} around the peak (red squares).
	These hashtags have neither the purely anticipatory nor the purely reactive behaviors illustrated above, and this may indicate a mix of exogenous and endogenous factors building up collective attention to a peak intensity, as a specific day approaches, and then away from it as user attention shifts away. See for example the case of the hashtag \texttt{\#watchmen}, used to discuss a blockbuster movie. The peak occurs on the day of the movie release in theatres.	
	\item Activity almost totally concentrated \textit{on the single day of the peak} (green rounded square). These hashtags correspond to transient collective attention associated with events that are highly discussed only while they happen, such as the $2009$ State of The Union address (\texttt{\#nsotu}), or the transient large-scale malfunctions of widely used Google services (\texttt{\#gfail}).
\end{itemize}
These patterns are somehow expected, in the sense that these are the only possibilities for the coarse-grained temporal profile of a hashtag with a popularity peak. However, the existence of well defined hashtag clusters, as well as their stability, are far from trivial and indicate that coarse graining the temporal dynamics of collective attention as shown here can expose robust indicators of the social semantics associated with hashtags. The presence of clearly separated clusters may also be deeply linked to the diverse nature of the mechanisms driving popularity in online social systems. Details on the usage and origin of the hashtags shown in Fig.~\ref{simplex} are available in Appendix~\ref{appendix-usage}.

%
%

\subsection{Social Semantics of Classes}

\begin{figure}[t]
\centering
\includegraphics[width=0.95\columnwidth]{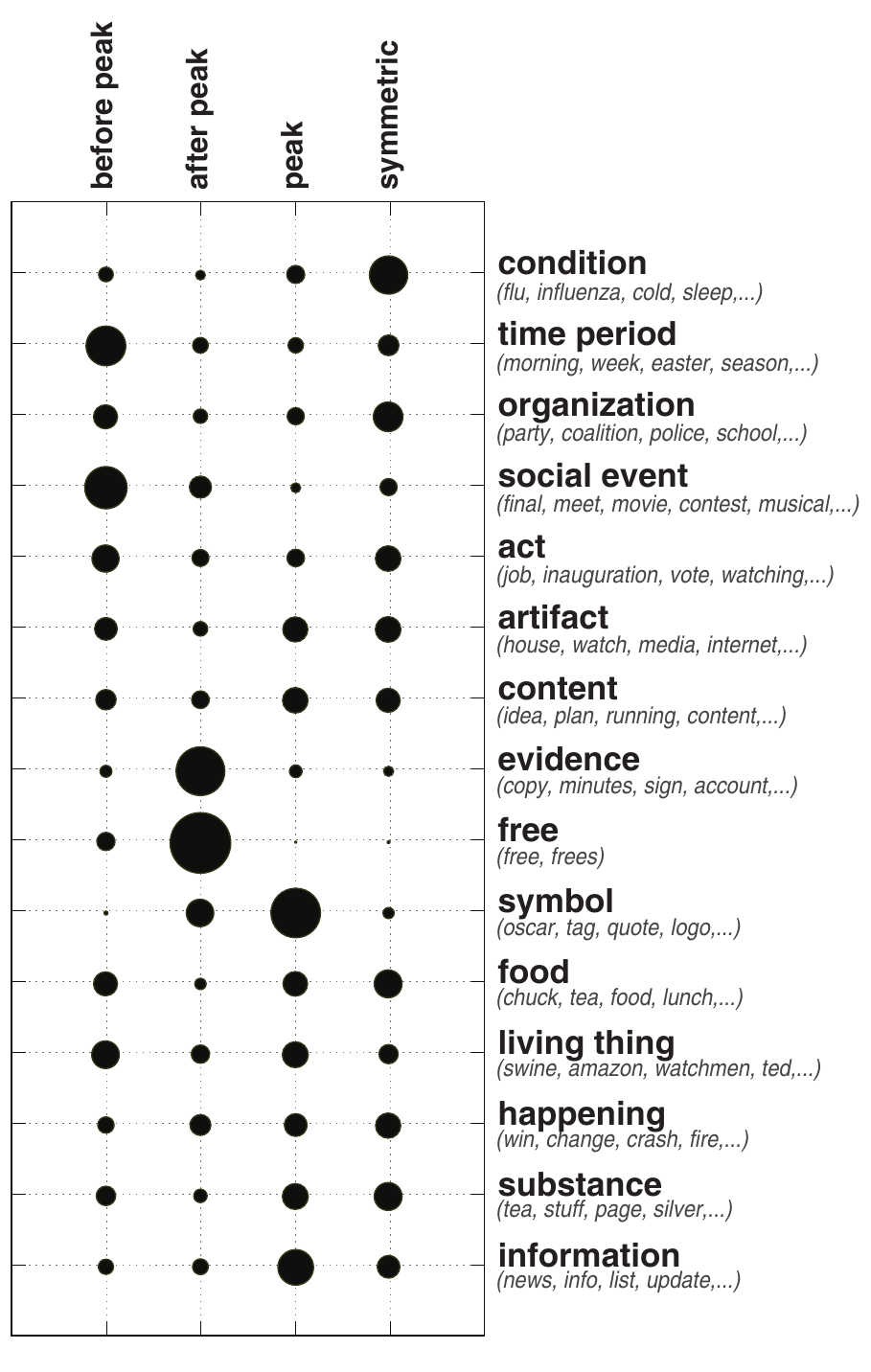}
\caption{
Semantic makeup of the hashtag classes: columns represent peak types and rows correspond to topics, i.e., concepts in the WordNet semantic lexicon. The radius of a circle is proportional to the average normalized frequency of the topic in the corresponding hashtag class. The displayed topics represent the most frequently observed generic concepts. Sample terms subsumed by them are reported in parenthesis.
\label{semantic} }
\end{figure}

The examples discussed above, such as those of Fig.~\ref{activity}, point to important differences in the social semantics of the different classes of hashtags. In order to shed light on this aspect, we systematically analyze the content of the tweets associated with each group of hashtags, using the semantic grounding described in Section~\ref{semanticanalysis}. WordNet provides hierarchical structures of concepts that can be made into a single directed acyclic graph by adding a root ``entity'' node as parent of the WordNet taxonomies. Thus, Wordnet can be used to coarse-grain the semantics of the looked-up terms by focusing on a given (high enough) level of the subsumption hierarchy.
Our interest here is to provide a semantic fingerprint of the content associated with the different hashtag classes, in order to expose differences in their social semantics. The concepts at depth $4$ of the WordNet hierarchy were identified as appropriate for this purpose, as that hierarchical level provides a good enough semantic diversity while featuring a small number of generic subsuming categories. We restricted our analysis to the concepts at depth $4$ that occur most frequently in the text associated with the hashtags under study: the right-hand side of Fig.~\ref{semantic} lists the $15$ selected WordNet concepts, together with sample terms that are subsumed by them.

To expose the semantic differences between hashtag classes we proceed as follows: For each hashtag we compute a normalized feature vector of the frequencies of occurrence of the selected WordNet concepts. We then average this vector over all hashtag belonging to a given class and obtain the class feature vectors of Fig.~\ref{semantic}, 
where the radius of discs is proportional to the normalized frequency of the corresponding concept in a given class of hashtags.
Clearly, different dynamic classes correspond to different semantics of the corresponding tweets.
The content of hashtags with activity concentrated before the peak has a stronger prevalence of concepts like ``social events'' and ``time period'' (e.g., \texttt{easter}), consistent with the social anticipation of a known event.
Conversely, hashtags whose activity is concentrated after the peak, usually associated to unexpected events, include several marketing campaigns such as \texttt{macheist}, and this is reflected in the prevalence of concepts like ``free'' and ``evidence''. Tags with the activity concentrated mostly on the peak day correspond to events that attract the users' attention for short periods of time, such as sport events and media events (e.g., concepts associated with \texttt{oscar}, subsumed by the ``symbol'' concept).
The detailed annotations of Appendix~\ref{appendix-usage} allow to make contact between specific hashtags or hashtag classes and the information of Figs.~\ref{simplex} and \ref{semantic}.
Notice that the observed selectivity between content and activity profiles may open the door to content tagging techniques based on popularity dynamics and on other behavioral cues.

\section{Information Spreading}
\label{sec:spreading}

Having identified classes of popular hashtags that differ in activity profiles and semantics, we now turn to investigating whether such classes are also associated with distinct patterns of information propagation.
Similarly to the approach of Ref.~\cite{crane08}, we regard information spreading as an epidemic process, where the behavior of using a given hashtag spreads from one user to another. The relevant social network for this epidemic process is Twitter's \textit{follower network}: whenever a user posts a given hashtag, her followers are exposed to the hashtag and can decide to adopt it in turn. Of course, users can also start using the hashtag spontaneously, as a result of exposure to external events (elections, sport matches, disasters, etc.) or to exogenous information sources.

\subsection{Basic Features}

The first feature we analyze is the fraction of retweets to total tweets in the messages associated with each hashtag under study. Retweets are forwarding actions in which a tweet from a followed user is delivered to one's followers together with a reference to the source. Because of their nature, retweets have been investigated as a mechanism for information diffusion in Twitter~\cite{galuba10}. 
The fraction of retweets is an indicator of how many (forwarded) copies are present in the tweets associated with a hashtag, and provides information on the spreading attitude of the corresponding topic. Retweets were identified both by checking for an initial ``RT'' marker or through tweet metadata. The top-left panel of Figure~\ref{epidemic} reports the fraction of retweets for the four hashtag classes. A box plot is used to provide information on the dispersion of parameter values inside each hashtag class. Hashtags with the activity distributed symmetrically around the peak or concentrated at the peak day have a higher fraction of retweets.
This supports the idea that those hashtags are associated with a higher level of endogenous activity, similarly to what happens for some YouTube videos~\cite{crane08}. Conversely, hashtags characterized by activity before the peak are associated to anticipatory behaviors and appear less prone to viral spreading.

The box-plot in the top-right panel of Fig.~\ref{epidemic} reports the fraction $\gamma$ of users who adopt the hashtag when none of the users they follow have used it before.
In other words, $\gamma$ estimates the fraction of ``seeders'' that inject the information related to the hashtag into the social network. Although the level of heterogeneity inside the four groups is high, we see that the hashtags with activity concentrated after the peak tend to have more seeders. 
This indicates that the propagation is probably fueled by exogenous factors, such as publicity campaigns or mass media communication. A further corroboration is provided by the semantic analysis of Fig.~\ref{semantic}, as these hashtags contain concepts such as ``sign'' (sign-up for a service) , ``account'' (create an account) or ``free'' that are usually associated with commercial campaigns that are heavily diffused in traditional media.

\subsection{Epidemic Parameters}

\begin{figure}
\centering
\includegraphics[width=0.9\columnwidth]{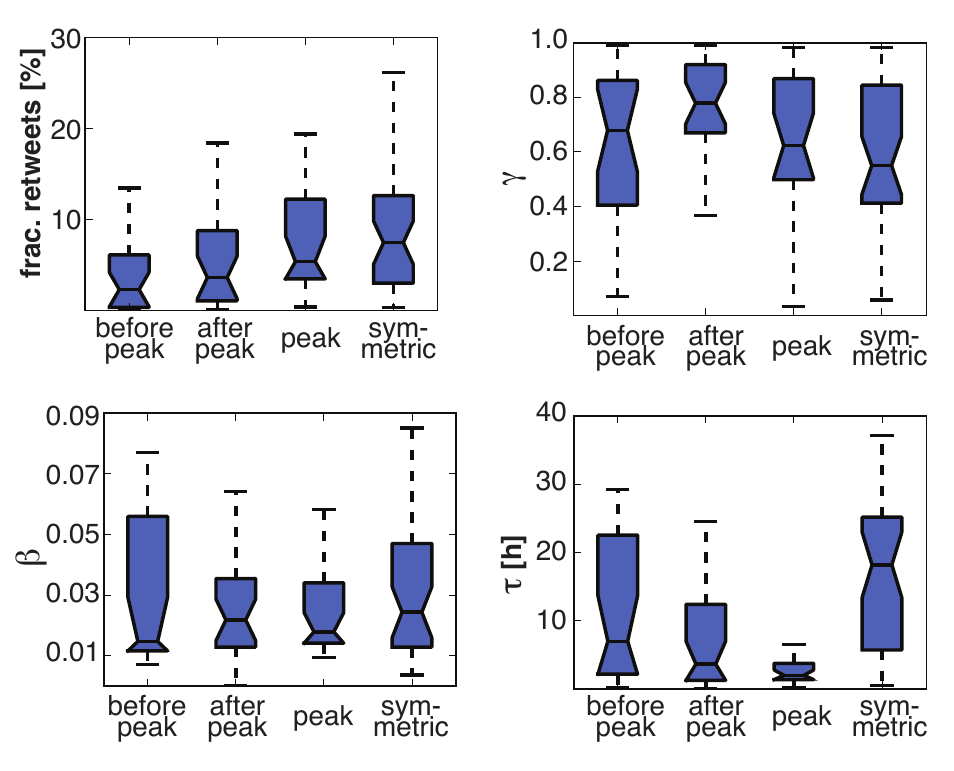}
\caption{
Parameters controlling the spreading of hashtags, broken down by hashtag class. Top left: fraction of retweets to regular tweets. Top right: fraction of seeders $\gamma$. Bottom left: fraction $\beta$ of followers that adopt the hashtag after seeing it. Bottom right: average time $\tau$ between the first tweet with the hashtag and the last one.
\label{epidemic}
}
\end{figure}

The box-plot in the bottom-left panel of Fig.~\ref{epidemic} reports the average fraction $\beta$ of a user's followers who adopt the hashtag after he or she has posted a tweet containing it. In modeling epidemic processes, $\beta$ is a measure of infectiousness. In this context, it bears information about the capacity of a behavior or meme to propagate from a user to her followers. The box-plot shows that $\beta$ does not depend strongly on the hashtag class and its median value is about $0.02$. 
This might suggest the existence of a generic mechanism controlling the propagation of the information over the Twitter social network independently of the content or popularity profile of the hashtags. 
The estimation of both $\gamma$ and $\beta$ depends on the sampling of the social network at hand. However, an analysis made using sub-samplings of the follower network obtained by cutting edges has showed that $\beta$ is relatively stable to the level of sampling, while $\gamma$ is more sensitive. Nevertheless, since our sampling of the network is fixed it is legitimate to compare the results obtained for different hashtags even in the case of $\gamma$.
     
Finally, in the bottom-right panel of Fig.~\ref{epidemic} we report the average time $\tau$, in hours, between the first tweet and the last tweet with the same hashtag posted by each user (we set $\tau = 0$ for those users who post the hashtag only once). That is, $\tau$ indicates the time during which users are likely to spread their use of the hashtag to followers.
The four hashtags classes display similar values of $\tau$ except for the case of activity concentrated on the peak day. In that case, hashtags have the lowest $\tau$ value, since activity is concentrated in a small period of time corresponding, for example, to a short-term disruption of online services.

\section{Discussion}
\label{sec:discussion}

In summary, we performed an extensive analysis of the Twitter hashtags that exhibit a popularity peak. Previous work found that popularity peaks in online systems can be clustered in a few prototypical classes according to the temporal features of their popularity dynamics.
Here we introduce a simple way of coarse-graining the temporal usage patterns of hashtags that exposes discrete dynamical classes. The clusters we find correspond to the four possible ways of distributing the hashtag activity with respect to the day of peak usage. Clusters are well defined and the classification of hashtags is stable with respect to small perturbations.
We ground in a semantic lexicon the contents of tweets associated with popular hashtags, and find insightful correlations between the class a hashtag belongs to and the (social) semantics of the associated content. 
In particular, hashtags that are mostly active before reaching a peak usually deal with scheduled social events or specific moments in time, indicating an anticipatory collective behavior. Hashtag with symmetric activity patterns across the peak seem to be associated with endogenous propagation over the social network. Hashtags that only exhibit a tail of activity after the peak correspond to unexpected events or exogenous driving.

Furthermore, we measure standard parameters of epidemic propagation over the on-line social network and relate these parameter values to the different hashtag classes, to unveil patterns of injection or propagation of information. The balance between internal propagation (endogenous) and external injection of information was assumed so far to be the main explanation for the existence of different clusters of online popular events. Our results indicate that the content type is also very important.
For instance, the hashtags used to discuss the ``swine flu'' pandemic (top of Fig.~{simplex}) or a popular event such as the Oscars ceremony (bottom-left of the simplex) show markedly different popularity profiles despite the fact that both attract a high level of attention from the media. Both hashtags display high levels of external seeding, as well as relatively low levels of endogenous propagation. Thus, the different social semantics of these hashtags is likely the cause underlying the observed differences in activity dynamics.


We remark that a robust classification into dynamical classes of user attention was obtained by using very simple parameters computed on time series of daily popularity. Contrary to other methods, which require the estimation of power-law exponents for popularity growth, or the computation of expensive correlations between high-resolution activity time series, the parameters introduced here can be easily computed in a scalable way.
While they lack predictive power, as they need a record of past activity to be computed, they can support the discovery of specific behavioral patterns in large-scale records of user activity. The robustness of the proposed approach, if confirmed in other settings, could support implicit temporal tagging of the Twitter data stream, where -- for example -- anticipatory behavior associated to a given date points to that date as a focus of collective expectation. The specific semantics that can be linked to a given temporal profile may be used to mine collective attention in order to construct implicit annotations of timelines on the basis of social media streams.
Of course, this requires an extensive work of validation that falls outside the scope of the present work. Progress in this direction will requires more refined content analysis by means of natural language processing and sentiment analysis, as well as validation in user studies or crowd-sourced settings.

\section{Acknowledgments}
The authors thank the PIs of the Truthy project, Fil Menczer, Alessandro Flammini, Johan Bollen and Alessandro Vespignani,  for their support and for many inspiring discussions. 
CC and JL thank Andr\'{e} Panisson for interesting discussions and technical help.
CC thanks Yamir Moreno for stimulating discussion.
This work was carried out while JL was at the ISI Foundation with support from the Leonardo da Vinci Scholarship.
JL acknowledges support from the Spanish Ministry of Science through the project TIN2009-14560-C03-01.
JJR acknowledges support from the JAE program of the CSIC and from the Spanish Ministry of Science (MICINN) through the project MODASS (FIS2011-24785).
CC acknowledges support from the Lagrange Project funded by the CRT Foundation and from the Q-ARACNE project funded by the Fondazione Compagnia di San Paolo.



\onecolumn
\appendix
\section{Hashtag Usage}
\label{appendix-usage}

\scriptsize
\begin{longtable}{p{70pt}p{100pt}p{220pt}}
\hline
\textbf{hashtag name} & \textbf{event type} & \textbf{description} \\
\hline
& \\
\multicolumn{3}{l}{activity before peak} \\
\hline
advertising & twitter game & shorty awards for advertisements \\
apps & twitter game & shorty awards for applications \\
asot400 & holiday/honor & event for the 400th episode of Armin van Buuren's radio show \\
cparty & convention & technology festival and LAN Party in Brazil (campus party) \\
earthhour & awareness/charity & event against climate change (turning off the lights for one hour) \\
easter & holiday/honor & celebration of Eastern \\
entertainment & twitter game & shorty awards for entertainment \\
firstfollow & twitter application & relates to {\#}FollowFriday \\
macworld & convention & MacWorld conference {\&} expo \\
masters & sport & golf tournament (masters cup) \\
mrtweet & twitter application & introduction of a new Twitter service to find people \\
myfirstjob & twitter game & sharing of first job experiences \\
nfl & sport & Super Bowl: Cardinals vs. Steelers \\
oneword & twitter game & tweeting of a word that's in the mind of Twitter user \\
plurk & twitter application & integration of Plurk into Twitter (service similar to Twitter) \\
poynterday & holiday/honor & honoring of Dougie Poynter \\
rncchair & political & RNC chairmanship election \\
sxswi & convention & set of film, interactive and music festivals (South by Southwest) \\
teaparty & political & protests across the United States \\
therescue & awareness/charity & event from the organization ``invisible children'' against child soldiers in Northern Uganda \\
tweepme & twitter game & contest for the twitter application TweepMe \\
twestival & awareness/charity & charity event of cities to raise money for clean water \\
wbc & sport & Japan's World Baseball Classic \\
\hline

& \\
\multicolumn{3}{l}{activity after peak} \\
\hline
amazonfail & disruption & demonstration against the new ranking of books in Amazon \\
americanidol & media & television competition to find new singing talents \\
blogger & twitter application & introduction of a new Twitter directory (WeFollow) \\
bsg & media & finale of Battlestar Galactica \\
contest & marketing/contest & competition to win the album ``Cardinology'' from Ryan Adams \\
cricket & sport & cricket game: India vs. England \\
earthday & awareness/charity & celebration of the earth day \\
evernoteclarifigiveaway & marketing/contest & competition to win iPhone 3G cases \\
free & marketing/contest & see {\#}MacHeist \\
fridayfollow & twitter game & unusual tag for {\#}FollowFriday \\
g20 & political & G-20 summit \\
happy09 & holiday/honor & congratulations to New Year's Eve \\
hoppusday & holiday/honor & honoring of Mark Hoppus of the band Blink182 \\
inaug09 & political & inauguration of Barrack Obama \\
job & twitter application & see {\#}tweetmyjob \\
macheist & marketing/contest & offering of free DEVONthink licenses from the Website MacHeist\\
mix09 & convention & conference for web designers and developers \\
peace & disruption & call of people for peace in Gaza \\
safari4 & technic & beta release of the web browser Safari 4 \\
skittles & marketing/contest & competition from the brand Skittles (candies) \\
spectrial & political & conviction of the Pirate Bay founders \\
starwarsday & media & Star Wars day (every May 4) \\
tweetmyjobs & twitter application & Twitter service for sending job posts \\
unfollowfriday & twitter game & countermovement to {\#}FollowFriday \\
winnenden & disruption & school shooting at a school in Winnenden, Germany \\
yourtag & twitter application & see {\#}blogger \\
zombies & disruption & see {\#}blackout \\
\hline

& \\
\multicolumn{3}{l}{activity at peak} \\
\hline
3hotwords & twitter game & tweeting of three hot word that's in the mind of Twitter user \\
aprilfools & holiday/honor & celebration of the April Fools' Day \\
bachelor & media & discussion of the finale episode of the reality show \textit{The Bachelor} in the night before \\
blackout & disruption & electricity blackout in Sydney \\
budget & political & delivering of the budget statement in UK \\
crapnames forpubs & twitter game & tweeting of worst names for a pub \\
followme stephen & twitter game & call to Stephen Fry to follow him \\
gfail & disruption & gMail blackout \\
gmail & disruption & see {\#}gfail \\
googmayharm & disruption & Google bug: Google may harm your computer \\
grammys & media & music award \\
horadoplaneta & awareness/charity & see {\#}EarthHour \\
mikeyy & disruption & worm attack in Twitter \\
nerdpickup lines & twitter game & tweeting of phrases about computers, star wars, etc. \\
nfldraft & sport & people are giving advices for the NFL draft \\
nsotu & political & first state of the union of Barrack Obama \\
oscar & media & movie award \\
oscars & media & see {\#}oscar \\
oscarwildeday & twitter game & competition by tweeting the best Wildean remarks, pics, etc. (game from Stephen Fry) \\
schiphol & disruption & airline crash at Amsterdam's Schiphol airport \\
snowmageddon & disruption & storm in Washington \\
superads09 & sport & advertisments during the Super Bowl \\
superbowl & sport & championship game of the NFL \\
superbowlads & sport & see {\#}superads09 \\
\hline

& \\
\multicolumn{3}{l}{activity before and after peak} \\
\hline
25c3 & convention & conference organized by the Chaos Computer Club \\
brand & twitter game & shorty awards for brands \\
bushfires & disruption & bushfires in Australien \\
cebit & convention & computer expo (CeBIT) \\
ces & convention & see {\#}ces09 \\
ces09 & convention & trade show for technology \\
chuck & media & see {\#}SaveChuck \\
coalition & political & prime minister in Canada won the right to suspend the parliament \\
davos & political & annual meeting of global political and business elites \\
dbi & twitter application & douche bag index is used from TweetSum to rank your followers by relevance \\
design & twitter game & shorty awards for design \\
drupalcon & convention & event for DrupalCon developers (content management system) \\
geek & twitter application & see {\#}blogger \\
glmagic & marketing/contest & competition to win over {\$}6,000 in electronics (from HP) \\
google & disruption & see {\#}googlemayharm \\
h1n1 & disruption & see {\#}swineflu \\
hadopi & political & adoption of the HADOPI law of control and regulation of Internet access in France \\
house & media & unexpected suicide of Lawrence Kutner, one of the main characters in the series Dr. House \\
humor & twitter game & shorty awards for humor \\
ie6 & activism & campaign against the usage of the IE6 \\
iloveyou & twitter game & call to post \textit{I love you} in online social networks \\
inauguration & political & see {\#}inaug09 \\
influenza & disruption & see {\#}swineflu \\
leweb & convention & Internet conference in Paris (LeWeb) \\
phish & media & reunion show of the American rock band Phish (Mar 6-8th, 2009) \\
pman & activism & protests against Moldovas parliamentary elections \\
politics & twitter game & shorty awards for politics \\
ptavote & twitter game & PTAVote platinum Twitter award \\
rp09 & convention & conference about Web 2.0 (re:publica) \\
safari & technic & see {\#}safari4 \\
savechuck & activism & call to save the television program \textit{Chuck} \\
skype & technic & iPhone OS release including the integration of Skype \\
socialmedia & twitter application & see {\#}blogger \\
swineflu & disruption & spread of the 2009 H1N1 virus (swineflu) \\
sxsw & convention & see {\#}sxswi \\
ted & convention & conferences of luminary speakers  \\
toc & convention & conference for the publishing and tech industries (Feb 9-11th 2009) \\
tweetbomb & twitter game & suggestion to bomb a person (mostly celebraties) with tweets \\
w2e & convention & Web 2.0 expo \\
watchmen & media & release of the movie \textit{Watchmen} \\
web & twitter application & see {\#}blogger \\
\hline
\end{longtable}

\end{document}